\documentclass[aps,twocolumn,preprintnumbers,amsmath,amssymb,prb,superscriptaddress]{revtex4}
\usepackage[dvips]{epsfig}
\usepackage{subfigure}
\usepackage{graphicx}
\usepackage{longtable}
\usepackage[T1]{fontenc}
\usepackage[ansinew]{inputenc}
\usepackage{ulem}
\usepackage{color}

\begin{document}

\title{Adsorption induced pyramidal distortion of the tri-metallic nitride core inside the endohedral fullerene Sc$_3$N@C$_{80}$ on the Ag(111) surface}

\author{Johannes Seidel}
\email[]{jseidel@rhrk.uni-kl.de}
\affiliation{Department of Physics and Research Center OPTIMAS, University of Kaiserslautern, Erwin-Schroedinger-Strasse 46, 67663 Kaiserslautern, Germany}
\author{Leah L. Kelly}
\affiliation{Department of Physics and Research Center OPTIMAS, University of Kaiserslautern, Erwin-Schroedinger-Strasse 46, 67663 Kaiserslautern, Germany}
\author{Markus Franke}
\affiliation{Peter Gr\"unberg Institut (PGI-3), Forschungszentrum J\"ulich, 52425 J\"ulich, Germany}
\affiliation{J\"ulich-Aachen Research Alliance (JARA) -- Fundamentals of Future Information Technology, 52425 J\"ulich, Germany}
\author{Gerben van Straaten}
\affiliation{Peter Gr\"unberg Institut (PGI-3), Forschungszentrum J\"ulich, 52425 J\"ulich, Germany}
\affiliation{J\"ulich-Aachen Research Alliance (JARA) -- Fundamentals of Future Information Technology, 52425 J\"ulich, Germany}
\author{Christian Kumpf}
\affiliation{Peter Gr\"unberg Institut (PGI-3), Forschungszentrum J\"ulich, 52425 J\"ulich, Germany}
\affiliation{J\"ulich-Aachen Research Alliance (JARA) -- Fundamentals of Future Information Technology, 52425 J\"ulich, Germany}
\author{Mirko Cinchetti}
\affiliation{Experimentelle Physik VI, Technische Universit\"at Dortmund, 44221 Dortmund, Germany}
\author{Martin Aeschlimann}
\affiliation{Department of Physics and Research Center OPTIMAS, University of Kaiserslautern, Erwin-Schroedinger-Strasse 46, 67663 Kaiserslautern, Germany}
\author{Benjamin Stadtm\"uller}
\email[]{bstadtmueller@physik.uni-kl.de}
\affiliation{Department of Physics and Research Center OPTIMAS, University of Kaiserslautern, Erwin-Schroedinger-Strasse 46, 67663 Kaiserslautern, Germany}
\affiliation{Graduate School of Excellence Materials Science in Mainz, Erwin Schroedinger Strasse 46, 67663 Kaiserslautern, Germany}

\date{\today}

\begin{abstract}
Our ability to understand and tailor metal-organic interfaces is mandatory to functionalize organic complexes for next generation electronic and spintronic devices. For magnetic data storage applications, metal-carrying organic molecules, so called single molecular magnets (SMM) are of particular interest as they yield the possibility to store information on the molecular scale. In this work, we focus on the adsorption properties of the prototypical SMM Sc$_3$N@C$_{80}$ grown in a monolayer film on the Ag(111) substrate. We provide clear evidence of a pyramidal distortion of the otherwise planar Sc$_3$N core inside the carbon cage upon the adsorption on the Ag(111) surface. This adsorption induced structural change of the Sc$_3$N@C$_{80}$ molecule can be correlated to a charge transfer from the substrate into the lowest unoccupied molecular orbital of Sc$_3$N@C$_{80}$, which significantly alters the charge density of the fullerene core. Our comprehensive characterization of the Sc$_3$N@C$_{80}$-Ag(111) interface hence reveals an indirect coupling mechanism between the Sc$_3$N core of the fullerene molecule and the noble metal surface mediated via an interfacial charge transfer. Our work shows that such an indirect coupling between the encapsulated metal centers of SMM and metal surfaces can strongly affect the geometric structure of the metallic centers and thereby potentially also alters the magnetic properties of SMMs on surfaces. 
\end{abstract}

\pacs{...}
\maketitle

\section{Introduction}
One of the great goals for future information technology is to develop new concepts to improve the speed and density of data storage applications for next generation spintronic devices. A highly promising approach to reduce the size of the individual data storage units down to the nanometer scale is to encode the information into the spin system of metal-carrying molecules, so called single molecular magnets (SMMs) \cite{Goodwin.2017, Guo.2017}. These molecular complexes possess the ability to retain magnetic order even without magnetic field for a considerable time at low sample temperatures \cite{Bogani.2008, Layfield.2014, Kuch.2016} making them highly interesting materials for short time storage applications. 

In most SMMs, the metal core is encapsulated by an extensive organic ligand system shielding the spin centers from direct interaction with its surrounding \cite{Sessoli.2009}. This prevents a modification of the magnetic moments of the spin centers due to chemical bonding with an underlying surface and protects the spin system against external contaminations such as oxidation. In this regard, endohedral fullerenes \cite{Bethune.1993, Popov.2013} are ideal model systems for SMM as they have the capability not only to encapsulate single magnetic rare-earth atoms, but also small clusters of several metallic centers. This chemical flexibility allows one to tune the spin structure of the SMM according to the desired functionality. 

Motivated by this opportunity, countless investigations were devoted to study and understand the intrinsic magnetic properties of SMMs in bulk-like molecular films (Ref. \cite{CHEM_CHEM200900728} and references therein). Only recently, the focus shifted towards monolayer films of SMM and endohedral fullerenes on ferromagnetic and noble metal surfaces \cite{Treier.2009, Huang.2011,Westerstrom.2012, Hermanns.2013, Muthukumar.2013, Westerstrom.2015, Leigh.2007, Kostanyan.2017}. One of the most frequently studied endohedral fullerene complexes on surfaces are tri-metallic nitride fullerenes (Me$_3$N@C$_{80}$) consisting of a very robust C$_{80}$ carbon cage which encapsulates three metal (Me) atoms coordinated to a central nitrogen (N) atom. These fullerene based SMMs arrange in long range ordered structures on noble metal surfaces and reveal magnetic ordering at low temperatures in zero field. Moreover, on ferromagnetic surfaces, an indirect carbon-cage-mediated exchange coupling of the metallic spin centers and the magnetic moments of the surface atoms was reported leading to the existence of electronically inequivalent metal centers within the fullerene cage \cite{Hermanns.2013}. These results clearly suggest that the magnetic properties of the spin centers of fullerene based SMMs on surfaces are influenced by the chemical interaction occurring at the metal-SMM interface, despite their protection by the fullerene cage. 

For this reason, we turn to the prototypical metal-SMM interface formed between the SMM Sc$_3$N@C$_{80}$ and the noble metal surface Ag(111) and investigate adsorption induced modifications of the geometric and electronic properties of the SMM and its encapsulated metal centers. This member of the family of Me$_3$N@C$_{80}$ fullerene molecules is particular interesting since the wave function of its lowest unoccupied molecular orbital (LUMO) is located not only at the carbon cage as it is expected for most endohedral fullerene molecules, but also at the Sc$_3$N core \cite{Zhang.2014,Popov.2017}. This is due to the large electronegativity of the Sc atoms with respect to other metal center such as yttrium or lanthanide atoms which are frequently found in tri-metallic nitride fullerenes. In addition, the Sc$_3$N group is located in the center of the fullerene cage in the gas phase with all four atoms located in a single plane\cite{Stevenson.1999}. This planar geometry inside the carbon cage is sterically and energetically favored due to the small ionic radius of the Sc$^{{3+}}$ centers of the Sc$_3$N group \cite{Dunsch.2007,Stevenson.1999}. On a surface, these geometric and electronic properties of the Sc$_3$N core can be monitored to identify a potential modification of the fullerene core by the molecule-surface interaction.

For our comprehensive characterization of the interfacial properties, we employ low energy electron diffraction (LEED), photoelectron spectroscopy (PES) and the normal incidence X-ray standing wave (NIXSW) technique. Similar to other Me$_3$N@C$_{80}$ complexes on noble metal surfaces, Sc$_3$N@C$_{80}$ grows in islands of a long range ordered structure in the (sub-) monolayer coverage regime on Ag(111). We observe a clear pyramidal distortion of the otherwise planar Sc$_3$N core with two chemically inequivalent Sc species located below the N atom. This geometric modification of the Sc$_3$N@C$_{80}$ complex coincides with an interfacial charge transfer from the substrate into the Sc$_3$N core. We hence propose that the observed geometric modification of the Sc$_3$N core is the result of the modified charge density inside the fullerene cage and the increased ionic radius of the Sc centers. These findings clearly underline the important role of the molecule-surface bonding for the geometric and the electronic, and therefore also for the magnetic properties of SMMs on surfaces.

\section{Experimental Details}
\subsection{Sample Preparation}
All experiments and the sample preparation were carried out under ultra-high vacuum condition with a base pressure better than 5*10$^{-10} \,$mbar. The surface of the Ag(111) single crystal was prepared by repeated cycles of argon ion sputtering (E$_{\mathrm{Ion}}=0.5\,$keV $-$ $1.5\,$keV, I$_{\mathrm{sample}}=8.0\,\mu$A) and thermal annealing up to $730\,$K for $20\,$min. The surface cleanness was confirmed either by the existence and line shape of the Shockley surface state at the $\bar{\Gamma}$-point of the surface Brillouin zone in valence band photoemission or by the absence of characteristic core level signals of surface contaminations using soft X-ray photoemission.  The Sc$_3$N@C$_{80}$ molecules (supplied by \textit{SES Research}, purity of $97\,\%$) were deposited onto the clean surface at room temperature using a dedicated thermal evaporation cell. The sublimation temperature of the Sc$_3$N@C$_{80}$ molecules was $800\,$K, the molecular coverage was monitored by the deposition time and calibrated by core level spectroscopy of the C1s signal. We used typical evaporation rates of $\approx 0.05\,\frac{\mathrm{ML}}{\mathrm{min}}$. The molecular films were annealed at the multilayer desorption temperature of T$_{\mathrm{sample}}=720\,$K for $10\,$min. This ensures the formation of a homogenous molecular film without any molecules adsorption in the second molecular layer. The sample quality was confirmed by LEED and core level photoemission. All measurements were performed with liquid helium cooling and a sample temperature of T$_{\mathrm{sample}}=30\,$K. The sample temperature was estimated from the increase of the Bragg energy when cooling the sample from room temperature to low temperature and the temperature dependent linear thermal expansion of Ag \cite{Xie.1999}.

\subsection{Normal Incidence X-ray Standing Waves}
The normal incidence X-ray standing waves (NIXSW) experiments were carried out at the Hard X-Ray Photoemission (HAXPES) and X-ray standing wave (XSW) end station of beamline I09 at the Diamond Light source (Didcot, UK). This end station is equipped with a hemispherical electron analyzer (Scienta EW4000 HAXPES) which is mounted perpendicular to the incoming photon beam. The angular acceptance of the electron analyzer is $\pm30^\circ$ and the energy resolution is $150\,$meV using the analyzer setting of the core level spectroscopy and the NIXSW experiment (E$_{\mathrm{Pass}}=100\,$eV, d$_{\mathrm{Slit}}=0.5\,$mm). 

The NIXSW method allows to determine the vertical adsorption position of all chemically different atomic species within an adsorbate system above a single crystal substrate with very high precision ($<0.04\,$\AA). In the following, we briefly summarize the fundamental aspects of this method. A more detailed introduction can be found elsewhere \cite{Woodruff+05,Zegenhagen}. For a photon energy fulfilling the Bragg condition $\vec{H}=\vec{k}_{H}-\vec{k}_{0}$ for a Bragg reflection $\vec{H}=(hkl)$, an X-ray standing wave field is formed by the interference of the incoming $\vec{E}_{0}$ and the Bragg-reflected wave $\vec{E}_{H}$. Scanning the photon energy through the Bragg condition results in a shift of the phase $\nu$ of the relative complex amplitude of the incoming and Bragg reflected wave by $\pi$. As a consequence, the standing wave field shifts by half a lattice spacing $d_{hkl}$ in the direction perpendicular to the Bragg planes. This changes the photon density at any specific position $z$ above the surface as a function of the photon energy. If an atom is located at this position $z$, the changes in its X-ray absorption yield during this scan can be monitored by recording the photoemission yield $I(E)$ of any of its core levels. The resulting experimental yield curve $I(E)$ can be modeled by\cite{Woodruff+05,Zegenhagen}
\begin{equation}  I(E)=1+R(E)+2\sqrt{R(E)}+F^H\cos{(\nu(E)-2\pi P^H)}, \end{equation}
where $R(E)$ is the X-ray reflectivity of the Bragg reflection with its complex amplitude $\sqrt{R(E)}$ and phase $\nu(E)$. The actual fit parameters are the coherent position P$ ^\mathrm{H}$ and the coherent fraction F$^\mathrm{H}$. P$^\mathrm{H}$ can be interpreted as the average vertical position D$^\mathrm{H}$ of an atomic species above the nearest lattice plane of the corresponding Bragg reflection H, which again is related to the true adsorption height $z$. F$^\mathrm{H}$ can be understood as a vertical ordering parameter with values between $0$ and $1$. F$^\mathrm{H}=0$ indicates complete vertical disorder of the emitting atomic species while for F$^\mathrm{H}=1$ all emitters are located at the same adsorption height corresponding to P$^\mathrm{H}$.

\section{Results}
\subsection{Lateral Structure}
\begin{figure}
	\centering
	\includegraphics[width=80mm]{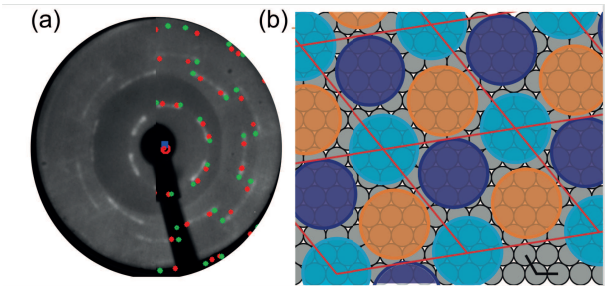} 
	\caption{(a) LEED image recorded for the monolayer film of Sc$_3$N@C$_{80}$ on Ag(111) at E$_{\mathrm{kin}}=32.5\,$eV and T$_{\mathrm{sample}}=30\,$K. The simulated diffraction pattern of our proposed structural models are superimposed onto the LEED data as red (structure R9) and green (structure R15) spots. (b) Structural model of the superstructure R9. The lattice of the R9 superstructure is shown in red, the positions and dimensions of the three Sc$_3$N@C$_{80}$ per unit cell are shown as colored circles.}
\label{Fig:Fig1}
\end{figure}

We have investigated the growth behavior and the lateral order of (sub-)monolayer films of Sc$_3$N@C$_{80}$ fullerenes on Ag(111) using LEED. At sub-monolayer coverages, the Sc$_3$N@C$_{80}$ molecules form islands of the long-range ordered monolayer structure. Increasing the coverage to one closed layer or changing the sample temperature from room temperature to $30\,$K does not change the lateral order of the molecular films. The corresponding LEED pattern of the monolayer structure is shown in Fig.~\ref{Fig:Fig1}(a). The diffraction spots are arranged in two concentric rings with $24$ spots in every ring. The molecular arrangement of the Sc$_3$N@C$_{80}$ on Ag(111) can be revealed by modelling the diffraction pattern. The best agreement between the experiment and the calculated diffraction pattern can be obtained for a superposition of two different superstructures with the superstructure matrices $\begin{pmatrix} 5.38 & -1.08\\ 1.08 & 6.47 \end{pmatrix}$ and $\begin{pmatrix} 5.81 & -0.36\\ 1.79 & 6.70 \end{pmatrix}$. The superstructure matrices express the relation between the molecular lattice and the grid of the Ag(111) surface. The simulated diffraction patterns of both superstructures are plotted onto the LEED data as red and green circles in the right part of Fig.~\ref{Fig:Fig1}(a). 

Both superstructures can be classified as non-commensurate point-on-line superstructures. This is consistent with the decreasing intensity of the diffraction spots with increasing diffraction order. The lattice parameters of both structures are equivalent to those of a commensurate $6\times 6$ superstructure that is rotated by $9 ^\circ$ (structure R9, red spots in Fig.~\ref{Fig:Fig1}(a)) and $15 ^\circ$ (structure R15, green spots in Fig.~\ref{Fig:Fig1}(a)) with respect to the $[\bar{1}10]$-direction of the substrate, respectively. For one molecule per unit cell, this would correspond to a nearest neighbor distance of $17.3\,$\AA\ which is significantly larger than the intermolecular distances reported for other Me$_3$N@C$_{80}$ molecules arranged in long range ordered superstructure on noble metals \cite{Hermanns.2013, Treier.2009}. Therefore, we propose a more densely packed molecular superstructure with three structurally inequivalent molecules per unit cell. A schematic real space model for such a structure is shown in Fig.~\ref{Fig:Fig1}(b) for the R9 superstructure. The three Me$_3$N@C$_{80}$ molecules are shown as circles of different color, the diameter of these circles corresponds to the diameter of a free Me$_3$N@C$_{80}$ molecule. From our model, we can estimate a nearest neighbor distance between two molecules of $\approx 10.0\,$\AA\ which is almost identical to the one reported for similar Me$_3$N@C$_{80}$ complexes on noble metals \cite{ Hermanns.2013, Treier.2009}. 

Note that the formation of large unit cells with three molecules per unit cell as well as the existence of more than one superstructure is not necessary surprising for endohedral fullerenes on noble metal surfaces. Dy$_3$N@C$_{80}$ on Cu(111) also forms ordered structures with two distinct azimuthal orientations and rather small domain sizes with less than $30$ molecules per domain \cite{Treier.2009}. Both observations suggest that several adsorption configurations of the fullerene cage (adsorption site and rotation of the carbon cage) on the surface lead to comparable adsorption energies for the different adsorption configurations and are hence energetically almost equivalent. We therefore propose that the existence of three inequivalent molecules per unit cell is due to three marginally different adsorption configurations of the carbon cage (different rotation, different part of the molecule pointing towards the Ag(111) surface) with almost identical adsorption energy.  

\subsection{Core Level Spectroscopy}

\begin{figure*}[t!]
	\centering
	\includegraphics[width=180mm]{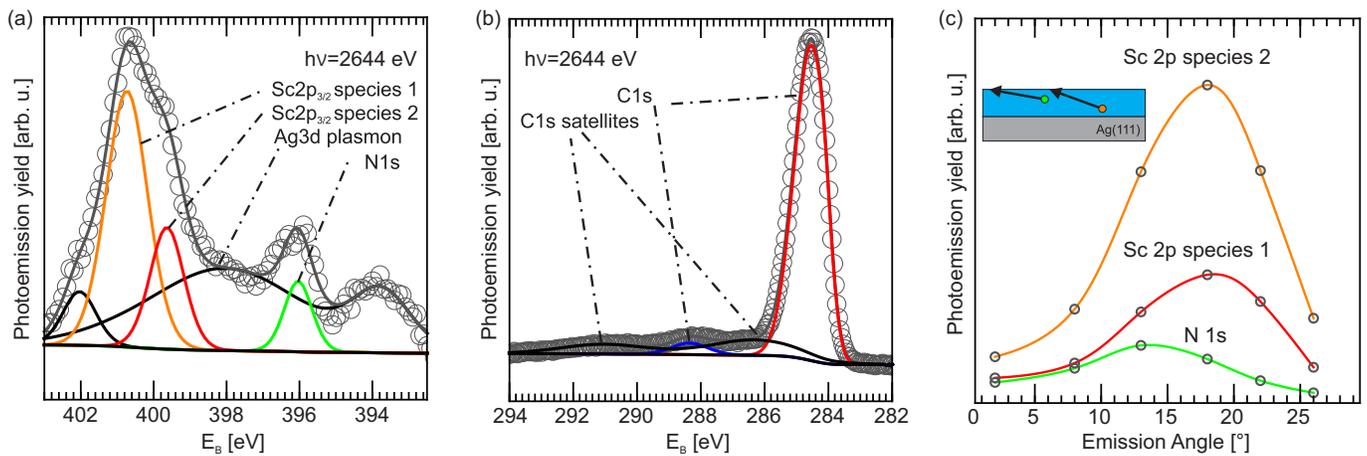}
	\caption{Core level spectra of N 1s and Sc 2p emission (a), and C 1s emission (b), as obtained from a monolayer film of Sc$_3$N@C$_{80}$/Ag(111). The energy scale was calibrated using the core level features of the Ag 3d bulk signal. The fitting model of all individual species are shown as Gaussian curves underneath the experimental data shown as data points. The signal of the photoemission background is described by three Gaussian curves to approximate the signature of the Ag 3d surface plasmon \cite{Stadler.2006,Mercurio.2010}. The core level data are recorded using hard x-ray photons ($\hbar\omega=2.644\,$keV) and at $30\,$K sample temperature. (c) Angle resolved core level yield of the N and Sc species of Sc$_3$N@C$_{80}$. The emission angles are measured with respect to the surface plane. The core level yield was extracted from the core level data using the dedicated fitting models shown in panel (a). Solid lines are drawn to guide the eye. The inset illustrates the relation between the escape depth of the photoelectrons and the emission angle.}
	\label{Fig:Fig2}
\end{figure*}

We now turn to the discussion of the core level spectroscopy data of a monolayer film of Sc$_3$N@C$_{80}$/Ag(111). The core level spectra of the N 1s, Sc 2p and C 1s emission lines were recorded in line with the NIXSW experiments, and therefore with hard X-ray photons of $\hbar\omega=2.644\,$keV. This photon energy is slightly higher than the Bragg energy in order to avoid any influence of the spatially varying X-ray standing wave field on the spectral line shape and intenstiy of the core level signals. Fig.~\ref{Fig:Fig2}(a) shows the N 1s and Sc 2p core level spectrum. It is dominated by three contributions: the Sc 2p$_{3/2}$ core levels of two different species at E$_\mathrm{B}=400.7\,$eV and E$_\mathrm{B}=399.6\,$eV, the N 1s core level at E$_\mathrm{B}=396.0\,$eV and the spectroscopic signal of Ag 3d surface plasmons\cite{Stadler.2006,Mercurio.2010} forming a photoemission background as well as the well-defined peak at E$_\mathrm{B}=393.7\,$eV. The spectral shape of the Ag 3d surface plasmons matches well with a recent NIXSW study by Stadler et al. \cite{Stadler.2006} and is consequently modeled by a similar fitting model with two Gaussian curves.  

The most striking observation in the Sc core level data is the presence of two chemically inequivalent species of Sc 2p$_{3/2}$. This has not been observed for bulk-like Sc$_3$N@C$_{80}$ films \cite{Stevenson.1999,Alvarez.2002} despite the existence of two geometrically (and hence chemically) different Sc species within the molecular cage. This finding points to a surface enhanced splitting of both inequivalent Sc centers within the endohedral fullerene on the Ag(111) surface. To quantify the spectral line shape of the Sc core levels, we employ the fitting model shown in Fig.~\ref{Fig:Fig2} (a). The two chemically inequivalent Sc signals are modeled by Gaussian curves of identical full width at half maximum (FWHM). We find a considerable chemical shift between both Sc species of $\Delta E_{\mathrm{B}}=1.0\,$eV and a relative ratio of both species of $\approx 2:1$, determined from the relative intensity of the corresponding photoemission lines. Note that an additional Gaussian curve at E$_\mathrm{B}=402.0\,$eV is necessary to describe the line shape of the Sc 2p$_{3/2}$ signal at its high binding energy side. This feature is attributed to a spectroscopic signal of the Ag(111) substrate since it reveals the characteristic signature of an Ag bulk species in the NIXSW analysis (i.e. same vertical position and same degree of vertical order). However, the microscopic origin of this additional Ag bulk species could not be determined so far.

The C1s photoemission spectrum is shown in Fig.~\ref{Fig:Fig2}(b). It is dominated by a main emission line at $E_{\mathrm{B,1}}=284.2\,$eV with a long flat electron energy loss tail at larger binding energies. Moreover, we find an additional peak at $E_{\mathrm{B,2}}=288.2\,$eV which could either be due to carbon atoms in direct contact with the Ag(111) surface or due to carbon atoms of the fullerene cage interacting directly with the Sc atoms inside the cage.  A quantitative analysis of the C 1s spectrum reveals a relative peak area of $3:72$ between the additional peak and the main line. An energetic shift of $4\,$eV favors the stronger interaction between C and Sc atoms, which are much closer than the Ag atoms. Although these results do not allow an unambiguous assignment of the second carbon peak to a particular carbon species within the molecule, this ratio favors the second model, i.e., the second carbon peak is most likely caused by carbon atoms in direct vicinity of the Sc centers.    

For the Sc 2p and the N 1s core level lines, additional angle resolved core level spectroscopy data were acquired using the angle-resolved detection mode of the hemispherical analyzer with an angular acceptance of $\pm 30^\circ$. The angle resolved core level data were recorded in a fixed experimental geometry with an almost normal incidence of light onto the sample surface and an angle of $90^\circ$ between analyzer axis and direction of incoming synchrotron beam. The angle-resolved core level spectra were analyzed using the fitting model discussed above (see Fig.~\ref{Fig:Fig2}(a)); the total yield of each chemically different species is shown in Fig.~\ref{Fig:Fig2}(c) as function of the emission angle. Starting from grazing emission ($0^\circ$), we first observe an increase in the photoemission yield of the N 1s signal with maximum intensity at an emission angle of $13^\circ$ followed by a decrease in photoemission intensity for larger angles. The initial increase in photoemission intensity towards larger emission angles with respect to the surface plain (i.e. towards normal emission geometry) is the result of the limited mean free path of photoelectrons in organic materials \cite{Graber.2011} and the different effective escape depth of photoelectrons for different emission angles as illustrated in the inset of Fig.~\ref{Fig:Fig2}(c). On the other hand, the decrease in photoemission intensity for larger angles is caused by the reduced sensitivity (reduced transmission) of the photoemission detector for large detection angles. A similar angle-dependent photoemission signal was found for both chemically inequivalent Sc 2p species with maximum photoemission yield  at a slightly larger angle of $18^\circ$, i.e., closer to normal emission. 
These findings can be understood when considering the limited mean free path of photoelectrons and the high surface sensitivity of photoelectron spectroscopy in grazing emission geometry. The atomic species closest to the last surface layer will reveal the strongest photoemission signal at small emission angle with respect to the surface plane. In case of Sc$_3$N@C$_{80}$, these are the N atoms. Therefore, our analysis of the angle-resolved core level spectroscopy data allows us to draw two conclusions: (i) The N atoms are located above the Sc atoms. (ii) Both Sc species are located at roughly the same adsorption height. These conclusions will be vital for the interpretation of our NIXSW data. 

\subsection{Vertical Adsorption Configuration}

\begin{figure*}[t!]
	\centering
	\includegraphics[width=100mm]{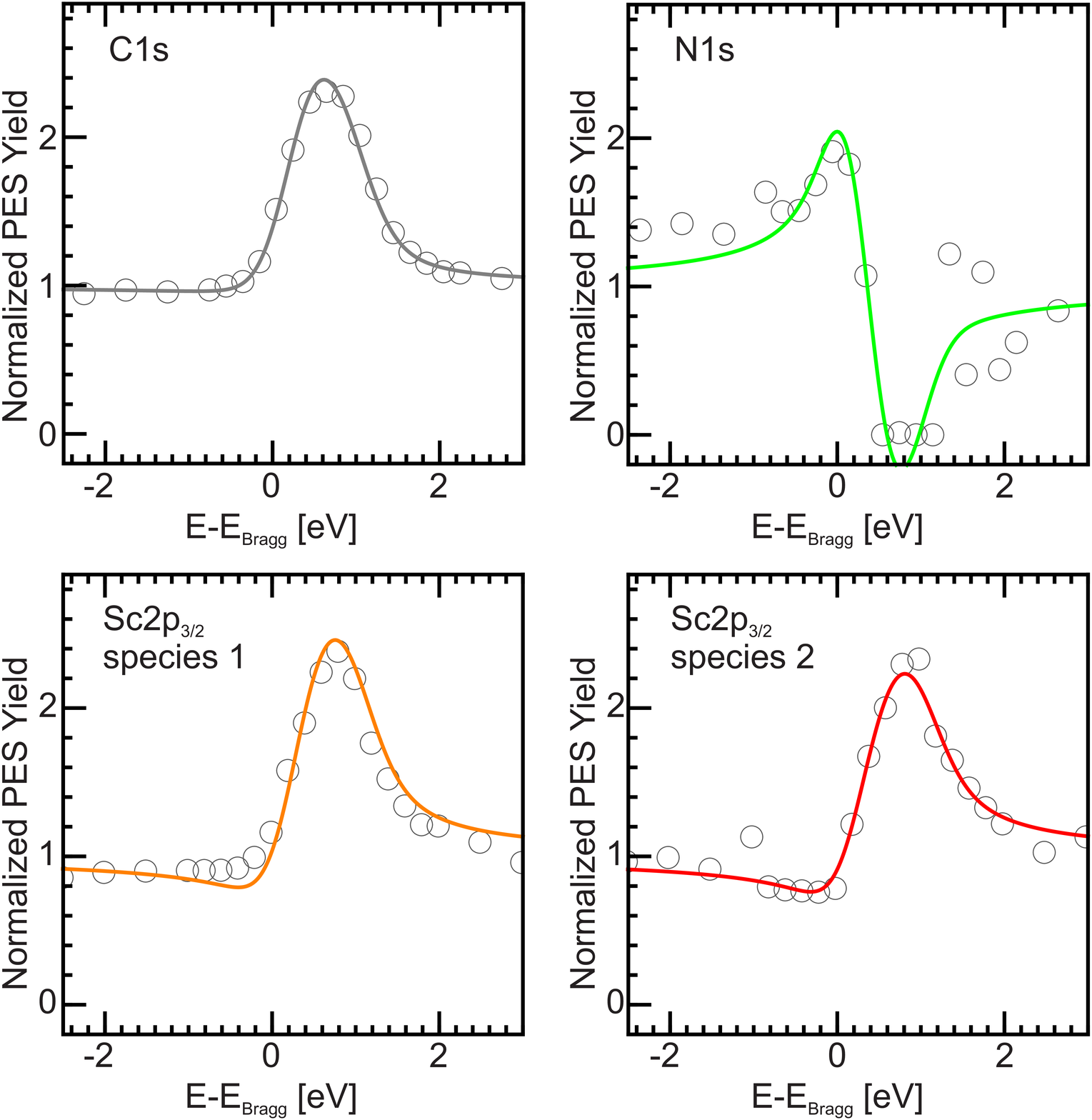}
	\caption{Exemplary partial yield curves of single NIXSW scans for all distinguished species. The yield curves were generated from the core level data using dedicated fitting models shown in Fig.~\ref{Fig:Fig2}. The solid lines represent the best fit to the data obtained with TORRICELLI \cite{toricelli,toricelli2}.}
	\label{Fig:Fig4}
\end{figure*}

\begin{figure*}[htb]
	\centering
	\includegraphics[width=160mm]{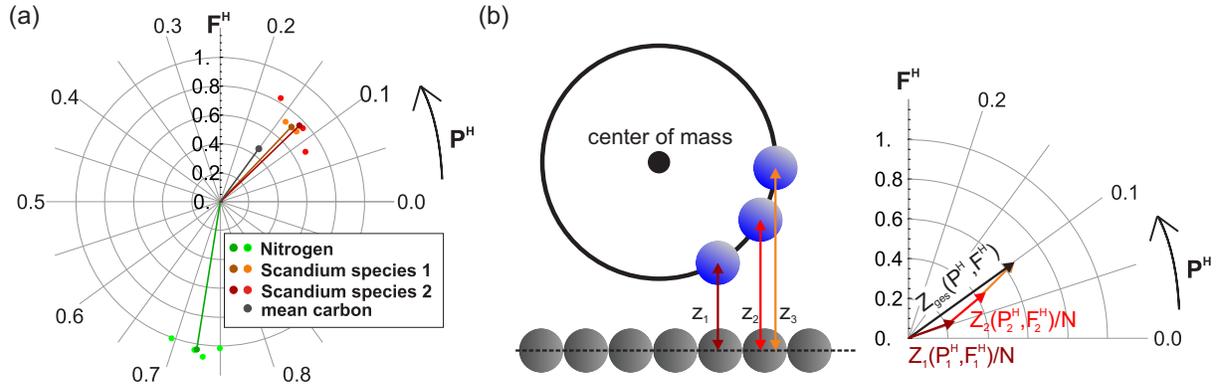}
	\caption{(a) Argand diagram illustrating the NIXSW fitting results for each individual NIXSW scan (colored circles) and the mean values of all scans for each species (colored line). Coherent positions P$^H$ and fractions F$^H$ are represented by the polar angle and the vector length, respectively. The color code is identical to that used for the yield curves (Fig.\ \ref{Fig:Fig4}). (b) Schematic illustration of our model simulations to estimate P$^H$ and F$^H$ of the carbon cage by adding up the contributions of the individual carbon atoms of the C$_{80}$ cage in the Argand diagram. For clarity, we only show three arbitrarily chosen carbon atoms at three different adsorption heights.}
	\label{Fig:Fig5}
\end{figure*}

In this section, we focus on the results of the NIXSW experiments. Typical photoemission yield curves for each species are shown in Fig.~\ref{Fig:Fig4}. They were obtained by analysis of the N 1s, C 1s and Sc 2p core level signals using the fitting models discussed above. This is only possible since these models were developed for core level spectra recorded with a very similar photon energy. Hence, the corresponding spectra reveal an identical satellite structure and a comparabile background signal as the core level spectra recorded during of the NIXSW scans. The uncertainty of each point was estimated by a Monte Carlo error analysis implemented in CasaXPS, \cite{MBW+13,toricelli,toricelli2} the software we used for fitting the core level spectra. They are usually smaller than $10\,\%$ and are omitted in Fig.~\ref{Fig:Fig4} for clarity. The yield curves were analyzed with the NIXSW analysis software TORRICELLI \cite{toricelli,toricelli2} resulting in the fitting parameters coherent position P$^\mathrm{H}$ and coherent fraction F$^\mathrm{H}$. During the experiment, NIXSW scans have been recorded for each species on several positions of the sample. The fitting results of all individual scans are shown as colored data points in the Argand diagram in Fig.~\ref{Fig:Fig5}(a), the averaged fitting parameters are summarized in table~\ref{Tab:Tab1}.

We start our discussion with the carbon species. We find an average coherent fraction of F$^\mathrm{H}=0.42$ and an average coherent position of P$^\mathrm{H}=0.15$. The low coherent fraction for the total carbon yield is not surprising when considering the 3D nature of the carbon cage and the large number of carbon atoms of the C$_{80}$ cage.  It can even be estimated for an undistorted Sc$_3$N@C$_{80}$ molecule by averaging the vertical adsorption position of the individual carbon atoms of the C$_{80}$ cage in the Argand diagram  \cite{Woodruff+98,Stadtmuller.2014}. This is illustrated in Fig.~\ref{Fig:Fig5}(b) for a small number of (carbon) atoms. The individual contribution of each carbon atom to the overall NIXSW signal can be represented by a vector $\bold{Z}_j (\mathrm{P}^\mathrm{H}(z_j),\mathrm{F}^\mathrm{H})= \mathrm{F}^\mathrm{H} e^{2\pi i \mathrm{P}^\mathrm{H}(z_j)}$ in the Argand diagram; $z_j$ is the vertical position of the $j$th carbon atom as obtained from density functional theory \cite{g16} and F$^\mathrm{H}$ is the coherent fraction of a single atom, i.e., F$^\mathrm{H}=1$ . The coherent fraction F$^\mathrm{H}$ of the entire C$_{80}$ cage is obtained by the vector sum of all vectors $\bold{Z}_j (\mathrm{P}^\mathrm{H}(z_j), \mathrm{F}^\mathrm{H})$, normalized to the number of carbon atoms of the C$_{80}$ cage. Using this formalism, we find a coherent fraction of F$^\mathrm{H}=0.19$ which is even smaller compared to the experimental value of F$^\mathrm{H}=0.42$. At first glance, this discrepancy could be explained by a vertical distortion of the carbon cage of the Sc$_3$N@C$_{80}$ molecule upon its adsorption on the Ag(111) surface. However, a precise quantification of the vertical distortion is challenging. We only find smaller coherent fractions when considering a linear vertical stretching or compression of the C$_{80}$ cage in our model simulations. This suggests that the undistorted carbon cage shows the highest vertical order with respect to the standing waves field. Along these lines, we also expect only marginal and non-uniform vertical distortion of the carbon cage due to the rather large F$^\mathrm{H}$ obtained in our experiment.  

\begin{figure*}
	\centering
 \includegraphics[width=160mm]{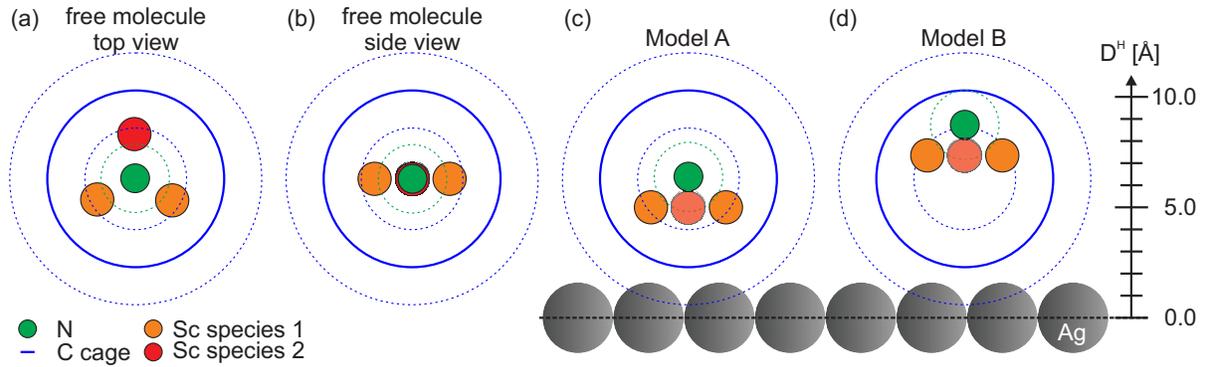}
\caption{Structural models of the Sc$_3$N@C$_{80}$ molecule prior (a),(b) and after (c),(d) its adsorption on the Ag(111) surface. In all cases, the carbon cage is illustrated schematically as a blue circle marking the center of mass of the carbon atoms of the fullerene cage. The nitrogen atom is shown as a green ball (atomic radius r$_{\mathrm{N, atomic}}=0.65\,$\AA). The two chemically different scandium species are shown as red and orange balls (ionic radius of the Sc$^{3+}$-ion, r$_{\mathrm{Sc, ionic}}=0.75\,$\AA\cite{Dunsch.2007}). The dashed green and blue lines mark the dimensions of the van der Waals radius of N and of the carbon cage. The structural models in panels (a) and (b) show a top view and a side view, respectively, onto the molecular geometry of a free Sc$_3$N@C$_{80}$ molecule in the gas phase. Panel (c) and (d) represent models of the two possible vertical adsorption configurations of Sc$_3$N@C$_{80}$/Ag(111) determined by NIXSW.}
 \label{Fig:Fig6}
 \end{figure*}

Using our model simulations for the undistorted Sc$_3$N@C$_{80}$ molecule, we only find one physically meaningful vertical adsorption position of the carbon cage above the surface. It is schematically illustrated in our vertical adsorption model in Fig.~\ref{Fig:Fig6} where the positions of the center of mass of the individual carbon cage atoms are represented by a blue ring. The center of mass of the carbon cage is located at z$=(6.27\pm 0.12)$\AA\ above the Ag surface, the carbon atoms closest to the surface at z$=(2. 18\pm 0.12)$\AA. This agrees well with previous findings for other fullerenes on noble metal surfaces \cite{Xu.2012, Pinardi.2014}. The distance between the lower part of the carbon cage and the silver surface is significantly smaller than the sum of the van der Waals radii of carbon (indicated by the outer dotted blue circle in Fig.~\ref{Fig:Fig6}) and silver and even smaller than typical bonding distances between planar carbon based organic molecules and the silver surface \cite{Kroger.2012, Kroger.2010, Hauschild.2010}. These findings point to a chemical interaction between the fullerene molecule and the Ag(111) surface. 
\begin{table}
\begin{tabular}{c|c|c}
Species & P$^{\mathrm{H}}$ & F$^{\mathrm{H}}$ \\
\hline
C 1s & 0.15 $\pm$ 0.05 & 0.42 $\pm$ 0.14 \\
N 1s & 0.72 $\pm$ 0.02 & 1.00 $\pm$ 0.03 \\
Sc 2p$_{\mathrm{species1}}$ & 0.13 $\pm$ 0.01 & 0.71 $\pm$ 0.01 \\
Sc 2p$_{\mathrm{species2}}$ & 0.12 $\pm$ 0.03 & 0.76 $\pm$ 0.05 \\

\end{tabular}
\caption{Averages fitting results obtained by NIXSW for a monolayer film of Sc$_3$N@C$_{80}$ on Ag(111) at T$_{\mathrm{sample}}=30\,$K.}
\label{Tab:Tab1}
\end{table}

For the N atoms, we observe an average coherent fraction of F$^\mathrm{H}\approx1$ pointing to a single adsorption height of all N atoms within the carbon cage despite the existence of three inequivalent molecules per unit cell. 
In contrast to the high F$^\mathrm{H}$ obtained for N1s we find coherent fractions for both Sc species of  F$^\mathrm{H}_\mathrm{species1}=0.71$ and F$^\mathrm{H}_\mathrm{species2}=0.76$, pointing to a certain degree of vertical disorder in the adsorption height of both Sc species within the carbon cage. The latter could be attributed to the existence of three inequivalent molecules per unit cell and hence to the existence of different vertical positions of the Sc atoms for the individual molecules of the unit cell. This suggests that the Sc atoms are more sensitive to the local environment of the Sc$_3$N@C$_{80}$ molecules than the N atoms. The coherent position of both Sc species is almost identical for both chemically different Sc species and we find a value of P$^\mathrm{H}_\mathrm{species1/2}=0.13/0.12$. The latter values are clearly different compared to P$^\mathrm{H}$ of nitrogen pointing to different vertical positions of both elements within the carbon cage. 

When calculating adsorption heights for N and both Sc species from the coherent positions, one has to consider that P$^\mathrm{H}$ only determines the vertical position $D^\mathrm{H}$ of an atom with respect to the lattice plane below. The real adsorption height z of each chemically different species is calculated by
\begin{equation} \label{DHcal} \mathrm{z}=(n+ \mathrm{P}^\mathrm{H})\times \mathrm{d}_{\mathrm{hkl}}, \end{equation} 
where $n$ is the number of the lattice planes between the surface and the adsorbed atom above the surface. For planar molecules on surfaces, usually only one value of $n$ results in physically meaningful adsorption height and P$^\mathrm{H}$ can be directly converted into an adsorption height z. This is more challenging for non-planar 3D molecules such as Sc$_3$N@C$_{80}$, which are significantly larger than the distance between two neighboring lattice planes of the Ag(111) surface (d$_{\mathrm{hkl}}=2.35\,$\AA\ at $30\,$K). However, in the following we show that we can still unambiguously reveal the most significantly adsorption induced modifications of the Sc$_3$N@C$_{80}$ molecule and can propose a vertical adsorption configuration for the Sc$_3$N core within the C$_{80}$ fullerene cage on Ag(111). 

Prior to the adsorption on Ag(111), the tri-metallic nitrite group is found in the center of mass of the almost spherical carbon cage with all four atoms located in one plane. This is illustrated by top and side views shown in Fig.~\ref{Fig:Fig6}(a),(b) for the free molecule. As already pointed out above, the experimentally obtained coherent position P$^\mathrm{H}$  of the carbon cage can only be converted into one meaningful vertical position of the carbon cage with its center of mass located at z$=(6.27\pm 0.12)$\AA\ above the Ag surface. 
In contrast, we obtain several possible vertical adsorption positions for the nitrogen and scandium atoms within the carbon cage. Two vertical adsorption models consistent with our angle resolved core level data (both Sc species are located below the N atom, see above) are shown in Fig.~\ref{Fig:Fig6}(c) and (d). The corresponding vertical positions of all chemically different species of both models A and B are summarized in table~\ref{Tab:Tab2}. However, only model A (Fig.~\ref{Fig:Fig6}(c))  fulfills the sterical requirements of the Sc$_3$N core within the carbon cage.  In model B, the position of the N atom is too close to the carbon atoms of the C$_{80}$ cage leading to an overlap of the non-interacting van der Waals radii of N and C (dashed green and blue circles). Therefore, we will only continue our discussion with model A. In this geometry, the N atoms are located $0.13\,$\AA\ above the center of mass of the carbon cage and are hence only slightly displaced from their position in the free molecule, i.e., from the center of the molecule. This is perfectly in line with recent investigations of other endohedral fullerene molecules on noble metal surfaces, which reported adsorption configurations with the N atoms located approximately in the center of the carbon cage \cite{Huang.2011, Treier.2009}. On the other hand, the Sc atoms are found at an average adsorption height of $\approx 1.4\,$\AA\ below the N atom, i.e., in the lower half of the carbon cage leading to an overall pyramidal distortion of the Sc$_3$N core upon the adsorption of the Sc$_3$N@C$_{80}$ molecule on Ag(111). 
Such a transition from the planar geometry of the Sc$_3$N in the gas phase to a pyramidal distorted geometry on Ag(111) is highly intriguing when considering the geometries of tri-metallic nitride cores for different Me$_3$N@C$_{80}$ complexes\cite{Dunsch.2007}. A planar geometry is usually found for endohedral fullerene complexes containing metal centers with small ionic radius \cite{Stevenson.1999, Stevenson.2002,Olmstead.2000} (such as Sc or Lu), while complexes with larger ionic centers \cite{Stevenson.2004,Zuo.2007} (such as Tb or Gd) reveal a pyramidal distortion of the encaged tri-metallic nitride group. The transition from a planar to a pyramidal geometry of the Me$_3$N core was attributed to the different size of the ionic radii of the metallic centers and the resulting sterical requirements of the Me$_3$N group within the C$_{80}$ cage. In the following, we will discuss the origin of this phenomena.
\begin{table}
\begin{tabular}{c|c|c}

Species & model A [\AA ] & model B [\AA ] \\
\hline
C 1s & 6.27 $\pm$ 0.12 & 6.27 $\pm$ 0.12 \\
N 1s & 6.40 $\pm$ 0.04 & 8.75 $\pm$ 0.04 \\
Sc 2p$_\mathrm{species1}$ & 5.00 $\pm$ 0.07 & 7.35 $\pm$ 0.07 \\
Sc 2p$_\mathrm{species2}$ & 4.99 $\pm$ 0.02 & 7.34 $\pm$ 0.02 \\
\end{tabular}
\caption{Adsorption heights of all chemically different species of Sc$_3$N@C$_{80}$/Ag(111) for both adsorption geometry models A and B shown in Fig~\ref{Fig:Fig6}.}
\label{Tab:Tab2}
\end{table}

\subsection{Valence Band Structure}

\begin{figure}
	\centering
 \includegraphics[width=80mm]{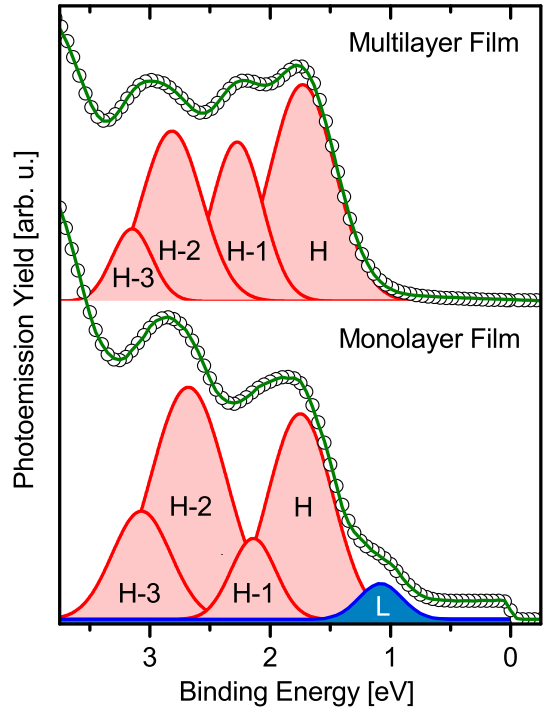}
\caption{Angle integrated valence band spectrum of the Sc$_3$N@C$_{80}$-Ag(111) interface (lower part) and of a Sc$_3$N@C$_{80}$ multilayer film on the Ag(111) surface (upper part). The spectral line shape was analyzed with a dedicated fitting model shown as red and blue Gaussian curves underneath the experimental data. The photoemission signal of the interface were recorded with a photon energy of $\hbar\omega=110\,$eV, the one of the multilayer film with $\hbar\omega=35\,$eV. Note that the valence band spectrum of the multilayer film was recorded at the NanoEsca endstation of the Elettra synchrotron (Trieste, Italy).}
 \label{Fig:Fig7}
 \end{figure}

In order to gain insight into the origin of the adsorption induced transition from a planar to a pyramidal geometry of the Sc$_3$N@C$_{80}$ core on the Ag(111) surface, we investigate the electronic valence band structure of the Sc$_3$N@C$_{80}$/Ag(111) interface using photoelectron spectroscopy. The PES spectrum of a monolayer film of Sc$_3$N@C$_{80}$ is shown in the lower part of Fig.~\ref{Fig:Fig7}, the one of a bulk-like multilayer film in the upper part for comparison. The multilayer spectrum is dominated by four rather narrow peaks that can be assigned to the highest occupied molecular orbital (HOMO, H in Fig.~\ref{Fig:Fig7}) at E$_\mathrm{B}=1.7\,$eV as well as to the energetically lower lying orbitals HOMO-1 (H-1), HOMO-2 (H-2) and HOMO-3 (H-3) at larger binding energies \cite{Alvarez.2002}. 

The PES spectrum of the Sc$_3$N@C$_{80}$ monolayer film closely resembles the spectrum of the multilayer film in the binding energy range between $1.5\,$ and $3.5\,$eV. All spectroscopic signatures of the four HOMO levels are found at almost identical binding energy positions as for the multilayer film and only reveal a marginal line shape broadening. The change in intensity is caused by the interaction of the molecule with the surface. In addition, the monolayer spectrum reveals a new state at E$_\mathrm{B}=1.0\,$eV (blue curve, L) which can be attributed to an adsorption induced state caused by the interaction between Sc$_3$N@C$_{80}$ and the silver surface. Most noticeable, the spectrum of the Sc$_3$N@C$_{80}$ monolayer film on Ag(111) is very similar to the one of a potassium (K) doped Sc$_3$N@C$_{80}$ multilayer film on an Au(110) single crystalline surface \cite{ Alvarez.2002}. In the latter case, a similar spectroscopic feature emerges at E$_\mathrm{B}\approx ($E$_\mathrm{HOMO}-0.5\,$eV) at moderate K doping concentration. This new state was assigned to the former LUMO which becomes occupied by the charge transfer from K to the Sc$_3$N@C$_{80}$ molecule.

Along these lines, we propose that the adsorption-induced state of the Sc$_3$N@C$_{80}$-Ag(111) interface can also be assigned to a LUMO derived interface state (LUMO-split-off-state) which becomes populated due to an interfacial charge transfer from the silver surface to the molecule. Adsorption induced charge transfer processes across molecule-metal interfaces are frequently observed for organic molecules on noble metal surfaces such as Ag(111) and Cu(111) and reflect the chemical character of the molecule-surface bonding in these adsorbate systems \cite{Willenbockel.2015, Gottfried.2015, Kilian.2008}. 

\section{Conclusion}

Our experimental findings for a long-range ordered monolayer film of the metal-carrying Sc$_3$N@C$_{80}$ endohedral fullerene on the Ag(111) surface reveal strong adsorption induced modifications of the geometric and electronic properties of the single molecular magnet on the surface. Using the NIXSW technique, we observe a pyramidal distortion of the usually planar Sc$_3$N core inside the C$_{80}$ cage. In addition, we find clear evidence for an interfacial charge transfer from the substrate into the LUMO of the Sc$_3$N@C$_{80}$ molecule. 
The pyramidal distortion of the Sc$_3$N core can either be attributed to a direct interaction between the Sc$_3$N core and the silver surface or to an indirect coupling mediated by the carbon cage. A direct interaction between the Sc$_3$N core and the surface is unlikely due to the robustness of the carbon cage which remains largely unaffected by the adsorption on the Ag(111) surface. Moreover, a direct coupling between the silver surface and the encaged Sc$_3$N core would most certainly not only reduce the adsorption height of the Sc atoms but also that of the N atom. However, the latter species is even pushed way from the interface (i.e., from the center of mass of the carbon cage) by the interaction of the endohedral fullerene with the Ag(111) surface. 

Instead, we propose that the pyramidal distortion of the core is caused by an indirect Sc$_3$N-surface interaction medicated by the carbon cage. The most striking evidence for such an indirect coupling is the interfacial charge transfer from the surface into the LUMO-derived split-off-state that was observed in our photoelectron spectroscopy data. As already discussed above, the wave function of the LUMO of the free Sc$_3$N@C$_{80}$ molecule is not only located at the carbon cage, but also at the Sc$_3$N core \cite{Shah.2015,Zhang.2014,Popov.2017}. In this way, the population of the LUMO significantly affects the electron density of the tri-metallic nitride core and hence modifies the electronic configuration of the encaged Sc$^{3+}$-ions. This additional charge density on the Sc sites leads to an increase of the ionic radii of the Sc atoms which alters the steric requirements of the Sc$_3$N group within the cage. As a result, the Sc$_3$N core is transformed from its planar gas phase geometry into the energetically more favorable pyramidal configuration upon its adsorption on Ag(111). A similar transition from a planar to a pyramidal configuration of the Me$_3$N core within a C$_{80}$ cage has already been observed for bulk-like films when increasing the size of the metal centers\cite{Dunsch.2007, Stevenson.1999, Stevenson.2002,Olmstead.2000, Stevenson.2004,Zuo.2007}. Additionally, the pyramidal distortion of the Sc$_3$N core could result in different positions of the Sc atoms with respect to the carbon cage and hence in the surface-enhanced difference in the chemical environment of the two chemically inequivalent Sc species observed in core level spectroscopy.

In conclusion, our comprehensive investigation of the geometric and electronic properties of the Sc$_3$N@C$_{80}$ - Ag(111) interface provides substantial evidence that the tri-metallic nitride core of endohedral fullerene molecules on surfaces is not completely shielded from the environment, i.e., the underlying surface. In particular, the chemical interaction between the C$_{80}$ carbon cage and the noble metal surface can result in an indirect interaction between the surface and the Sc$_3$N core via a carbon cage which can severely alter the geometric, electronic and potentially also the magnetic properties of the encapsulated metal centers. It leads to a pyramidal distortion of the otherwise planar Sc$_3$N core. Hence, this type of indirect coupling mechanism must be considered when designing metal carrying fullerene complexes as single molecular magnets with dedicated functionalities. This is particularly important in case the most frontier orbitals are located at the metallic spin-centers as demonstrated here for the endohedral fullerene Sc$_3$N@C$_{80}$ on Ag(111). On the other hand, the coupling might also provide an intriguing tool to manipulate the magnetic properties of the molecular units via spinterface effects \cite{Cinchetti.2017}. 

\section*{Acknowledgements}
The research leading to these results was financially supported by the Deutsche Forschungsgemeinschaft (DFG, SFB/TRR 88 "Cooperative Effects in Homo- and Heterometallic Complexes (3MET)" Project C5) and the Carl-Zeiss-Stiftung. B.S. thankfully acknowledges financial support from the Graduate School of Excellence MAINZ (Excellence Initiative DFG/GSC266). M.C. acknowledges  funding from the European Research Council (ERC) under the European Union's Horizon 2020 research and innovation programme (grant agreement n° 725767 - hyControl). Moreover, we thank Diamond Light Source for access to beamline I09 (through proposal SI13773-1). We are very grateful for the support by the beamline staff during the experiment, in particular by P. K. Thakur, D. A. Duncan, T.-L. Lee and D. McCue.


\end{document}